# Recent progress in low-energy electron elastic-collisions with multi-electron atoms and fullerene molecules

Alfred Z. Msezane and Zineb Felfli

Department of Physics and Center for Theoretical Studies of Physical Systems, Clark Atlanta University, Atlanta, Georgia 30314, USA

Abstract

We briefly review recent applications of the Regge pole analysis to low-energy $0.0 \leq E \leq 10.0$ eV electron elastic collisions with large multi-electron atoms and fullerene molecules. We then conclude with a demonstration of the sensitivity of the Regge pole-calculated Ramsauer-Townsend minima and shape resonances to the electronic structure and dynamics of the Bk and Cf actinide atoms, and their first time ever use as novel and rigorous validation of the recent experimental observation that identified Cf as a transitional element in the actinide series (A. Müller, et al., Nat. Commun. 12, 948 (2021))

Professor M. Ya. Amusia *taught us not to aim at merely reproducing experimental results but at understanding the fundamental physics*

## 1.0 Introduction

Progress towards the theoretical understanding of the fundamental mechanism underlying stable negative-ion formation in low-energy electron collisions with complex heavy multi-electron atoms and fullerene molecules has been very slow. This physical mechanism is of fundamental importance in physics and chemistry. More specifically, it has important implications for a wide range of applications, from catalysis to drug delivery and water purification. Unfortunately, the complexity of the interactions among electrons in heavy multi-electron atoms and fullerene molecules has for a long time made it virtually impossible to reliably predict the energetics of the electron binding and the properties of the resulting negative ions. A theoretical breakthrough was achieved in low-energy electron scattering from complex heavy multi-electron systems through our rigorous Regge pole method, wherein is embedded the electron-electron correlation effects and the core-polarization interaction, identified as the two crucial physical effects responsible for electron attachment resulting in stable negative-ion formation.

Consequently, the robust Regge pole method has allowed us to explore reliably for the first time ever negative-ion formation in complex heavy multi-electron systems such as the lanthanide and actinide atoms as well as the fullerene molecules through the electron elastic total cross sections (TCSs) calculation. Importantly, these yield directly the anionic binding energies (BEs), the shape resonances (SRs) and the Ramsauer-Townsend(R-T) minima. From the TCSs unambiguous and reliable ground, metastable and excited states negative-ion BEs of the formed anions during the collisions are extracted and compared with the measured and/or calculated electron affinities (EAs) of the atoms and fullerene molecules. The novelty and generality of the Regge pole approach is in the extraction of rigorous negative-ion BEs from the TCSs, without any assistance whatsoever from either experiment or any other theory. Whether the measured EAs are identified with the ground state anionic BEs or with the excited states anionic BEs of the formed negative ions during the collisions, the rigorous Regge pole-calculated BEs are available to guide measurements.

Essential to understanding chemical reactions involving negative ions are accurate and reliable atomic and molecular affinities [1]. Also, low-energy electron collisions, resulting in negative ion formation, provide a special insight into quantum dynamics [2]. Consequently, needed is the careful determination of the EAs, the Ramsauer-Townsend (R-T) effect, important *inter alia* for understanding sympathetic cooling and the production of cold molecules using natural fermions and SRs. Additionally, the EA provides a stringent test of theoretical calculations when their results are compared with those from reliable measurements. For ground states collisions the Regge pole-calculated negative ion BEs correspond to the challenging to calculate theoretically EAs, yielding outstanding agreement with the standard measured EAs for Au, Pt and the highly radioactive At atoms as well as for the $C_{60}$ and the $C_{70}$ fullerene molecules. For the fullerenes $C_{20}$ through $C_{92}$ our Regge pole-calculated ground-state anionic BEs have been found to match in general excellently the measured EAs [16, 17]. These results give great credence to the power and ability of the Regge pole method to produce unambiguous and reliable ground state anionic BEs of complex heavy systems through the TCSs calculation. Significantly, the Regge pole method achieves the remarkable feat without assistance from experiment or any other theory.

Unfortunately, for most of the lanthanide atoms, producing sufficient anions that can be used in photodetachment experiments is challenging [3]. Due to their radioactive nature the actinide atoms are difficult to handle experimentally. Thus the great need for reliable theoretical EAs to guide measurements. The EAs of atomic Au, Pt, and At have been measured [4-9], including those of the $C_{60}$ and $C_{70}$ fullerene molecules [10-14]. For the highly radioactive At atom various sophisticated theoretical calculations, including the Multiconfiguration Dirac Hartree-Fock (MCDHF) value [15] agree excellently with the measured EA [9]. Reference [9] employed Coupled-Cluster method, while Ref. [15] used MCDHF method. Furthermore, in [15] an extensive comparison among various sophisticated theoretical EAs has been carried out. For all these atoms the measured EAs matched excellently the Regge pole-calculated BEs of the anionic ground states of the formed negative ions during the collisions, see Table 3.1 for comparisons. Also, the measured EAs of the fullerenes $C_{20}$ through $C_{92}$ agree excellently with the Regge pole-calculated anionic ground states BEs [16, 17]. This gives great credence to our interpretation of the EAs of these complex systems, viz. as corresponding to the ground states BEs of the formed negative ions during the collisions.

Recently, the EAs of the highly radioactive actinide atoms Th[18] and U[19, 20] were measured as well. The experimentalists concluded that the EAs of both Th and U corresponded to the BEs of the weakly bound electron to the neutral atoms. For the Ti atom two measurements obtained the EAs as 0.377 eV[21] and 0.075 eV[22]. The former value is close to various theoretical calculations [23, 24], including the Regge pole-calculated BE of the second excited state, namely 0.281 eV[25]. However, the value of 0.075eV[22] is close to the Regge pole BE of the highest excited state of the formed Ti¯ anion, 0.0664 eV; its ground state BE is 2.42 eV [25]. The measured EA of Hf is 0.178 eV[26]. It is close to the Regge pole SR at 0.232 eV, the RCI EA of 0.114 eV[27] and the Regge pole second excited state anionic BE of 0.113 eV[28]. The Hf highest excited state BE is at 0.017 eV[29]. Indeed, here we are faced with the problem of interpretation of what is meant by the EA.

For the lanthanide atoms problems regarding what is meant by the EA have already been discussed [29, 30]. Briefly, for the Nd atom, there are two measured EA values, viz. 1.916 eV [31] and 0.0975 eV [32]. The value of [31] is close to the Regge pole ground state anionic BE value of 1.88 eV [33], while the EA of [32] is close to the RCI EA [34] and the Regge pole anionic BE of the highest excited state [33]. Similarly the measured EAs for the Eu atom are 0.116 eV[3] and 1.053 eV[35]. The former value agrees excellently with the Regge pole BE of the highest excited state, viz. 0.116 eV[33], see also Fig. 3.4 here, and with the RCI EA of 0.117 eV[36]. The EA of [35] agrees very well with the Regge pole-metastable anionic BE value of 1.08 eV[33]. For the large Tm atom the measured EA [37] is close to the Regge pole-metastable BE (1.02 eV) [33]. Clearly, the results here demonstrate the need for an unambiguous meaning of the EA. The crucial question also considered here is: does the EA of heavy multi-electron systems (atoms and fullerene molecules) correspond to the BE of the attached electron in the ground, metastable or excited state of the formed negative ion during the collision? Indeed, the meaning of the measured EAs of multi-electron atoms and fullerene molecules is also discussed here within the context of two prevailing viewpoints:

1) The first considers the EA to correspond to the electron BE in the ground state of the formed negative ion during collision; it is exemplified by the measured EAs of Au, Pt and At atoms and the fullerene molecules from $C_{20}$ through $C_{92;}$ and
2) The second view identifies the measured EA with the BE of electron attachment in an excited state of the formed anion. The measured EAs of Ti, Hf, lanthanide and actinide atoms provide representative examples of this viewpoint.

We conclude the paper with a demonstration of the first ever use of the Regge pole-calculated TCSs as probes of the electronic structures of the actinide atoms Bk and Cf to identify the transitional element in the actinide series [38].

## 2.0 Method of Calculation

In this paper we have used the rigorous Regge pole method to calculate the electron elastic TCSs. Regge poles, singularities of the S-matrix, rigorously define resonances [39, 40] and in the physical sheets of the complex plane they correspond to bound states [41]. In [42] it was confirmed that the Regge poles formed during low-energy electron elastic scattering become stable bound states. In the Regge pole, also known as the complex angular momentum (CAM), method the important and revealing energy-dependent Regge Trajectories are also calculated. Their effective use in low-energy electron scattering has been demonstrated in for example [33, 43]. The near-threshold electron–atom/fullerene collision TCS resulting in negative-ion formation as resonances is calculated using the Mulholland formula [44]. In the form below, the TCS fully embeds the essential electron-electron correlation effects [45, 46] (atomic units are used throughout):

$$\sigma_{tot}(E) = 4\pi k^{-2} \int_0^\infty \text{Re}[1 - S(\lambda)]\lambda d\lambda$$
$$-8\pi^2 k^{-2} \sum_n \text{Im} \frac{\lambda_n \rho_n}{1 + \exp(-2\pi i \lambda_n)} + I(E) \quad (1)$$

In Eq. (1) $S(\lambda)$ is the S-matrix, $k = \sqrt{2mE}$, $m$ being the mass and $E$ the impact energy, $\rho_n$ is the residue of the S-matrix at the $n^{th}$ pole, $\lambda_n$ and $I(E)$ contains the contributions from the integrals along the imaginary λ-axis (λ is the complex angular momentum); its contribution has been demonstrated to be negligible [33].

As in [47] here we consider the incident electron to interact with the complex heavy system without consideration of the complicated details of the electronic structure of the system itself. Therefore, within the Thomas-Fermi theory, Felfli et al [48] generated the robust Avdonina-Belov-Felfli (ABF) potential which embeds the vital core-polarization interaction

$$U(r) = -\frac{Z}{r(1+\alpha Z^{1/3}r)(1+\beta Z^{2/3}r^2)} \qquad (2)$$

In Eq. (2) $Z$ is the nuclear charge, $\alpha$ and $\beta$ are variation parameters. For small $r$, the potential describes Coulomb attraction between an electron and a nucleus, $U(r) \sim -Z/r$, while at large distances it has the appropriate asymptotic behavior, viz. $\sim -1/(\alpha\beta r^4)$ and accounts properly for the polarization interaction at low energies. For an electron, the source of the bound states giving rise to Regge Trajectories is the attractive Coulomb well it experiences near the nucleus. The addition of the centrifugal term to the well 'squeezes' these states into the continuum [49]. For larger complex angular momentum (CAM) $\lambda$ the effective potential develops a barrier. Consequently, a bound state crossing the threshold energy E = 0 in this region may become a long-lived metastable state or an excited state. As a result the highest "bound state" formed during the collision is identified with the highest excited state, here labeled as EXT-1. As E increases from zero, the second excited state may form with the anionic BE labeled, EXT-2. For the metastable states, similar labeling is used as MS-1, MS-2, etc. However, it should be noted here that the metastable states are labeled relative to the anionic ground state. Regge poles are generalized bound-states, namely solutions of the Schrödinger equation where the energy E is real, positive and $\lambda$ is complex. The CAM methods have the advantage in that the calculations are based on a rigorous definition of resonances, viz. as singularities of the S-matrix, see [49, 50].

The strength of this extensively studied potential [51, 52] lies in that it has five turning points and four poles connected by four cuts in the complex plane. The presence of the powers of Z as coefficients of $r$ and $r^2$ in Eq. (2) ensures that spherical and non-spherical atoms and fullerenes are correctly treated. Also appropriately treated are small and large systems. The effective potential $V(r) = U(r) + \lambda(\lambda+1)/2r^2$ is considered here as a continuous function of the variables $r$ and complex $\lambda$. The details of the numerical evaluations of the TCSs have been described in [46] and references therein; see also [53]. In the solution of the Schrödinger equation as described in [46], the parameters "α" and "β" of the potential, Eq. (2) are varied. With the optimal value of α = 0.2 the β-parameter is then varied carefully and when the dramatically sharp resonance appears in the TCS it is indicative of negative ion formation; this energy position matches the measured EAs of the atom/fullerene, for example Au or $C_{60}$ fullerene. This has been found to be the case in all the atoms and fullerenes we have investigated thus far.

## 3.0 Results
### 3.1 Cross Sections for the representative atom Au and fullerene molecule $C_{60}$

Figure 3.1, taken from Ref. [54] presents TCSs for atomic Au and fullerene molecule $C_{60}$. They typify the TCSs of complex heavy multi-electron atoms and fullerene molecules, respectively. Importantly, they are characterized by dramatically sharp resonances representing negative-ion formation in ground, metastable and excited anionic states, R-T minima and SRs. In both Figs. the red curves represent ground states electron TCSs while the green curves denote excited states TCSs. Here the ground states anionic BEs in both Au and $C_{60}$ appearing at the absolute R-T minima matched excellently the measured EAs, see Fig. 3.1 and Table 3.1 for comparisons with various measurements. In both systems, the ground states anionic BEs determine their EAs and not the excited anionic BEs (green curves). The data in Table 3.1 for Pt, At, Ti and Hf atoms as well as for the $C_{70}$ fullerene were extracted from similar curves as in Fig. 3.1. Notably, the TCSs for the atoms and fullerene molecules become more complicated as the systems considered become larger as exemplified by the actinide atoms in Fig. 3.7 and the fullerene molecules in [16].

The availability of excellent measured EAs for the Au and Pt atoms [4-8] and the $C_{60}$ fullerene molecule [10-12] allowed us to implement the rigorous Regge pole method to complex multi-electron atoms and fullerene molecules. Excellent agreement with the measurements was obtained for the Au, Pt and At atoms and the fullerene molecules $C_{60}$ and $C_{70}$ as demonstrated in Table 3.1. Thus the Regge pole-calculated ground state anionic BEs were benchmarked on the measured EAs of both Au and $C_{60}$. Subsequently, the Regge pole method was implemented in the calculation of low-energy electron elastic TCSs for various complex multi-electron atoms, including atomic Hf and Ti. The calculated ground states anionic BEs for the Au, Pt and At atoms matched excellently the measured EAs of these atoms. For the fullerene molecules $C_{20}$ through $C_{92}$ the obtained anionic ground states BEs [16, 17] agreed in general very well with the measured EAs. Indeed, the Regge pole method accomplished an unprecedented feat in the calculation of the challenging to calculate theoretically EAs of both multi-electron atoms and the fullerene molecules, from $C_{20}$ through $C_{92}$.

Table 3.1 already demonstrates the ambiguous meaning of the measured EAs in the large atoms Ti and Hf versus the meaning in the Au, Pt and At atoms. The interpretation of the EAs in the former atoms is that they correspond to the anionic BEs of excited states, while in the latter atoms the EAs are identified with the Regge pole BEs of the formed anions in their ground states

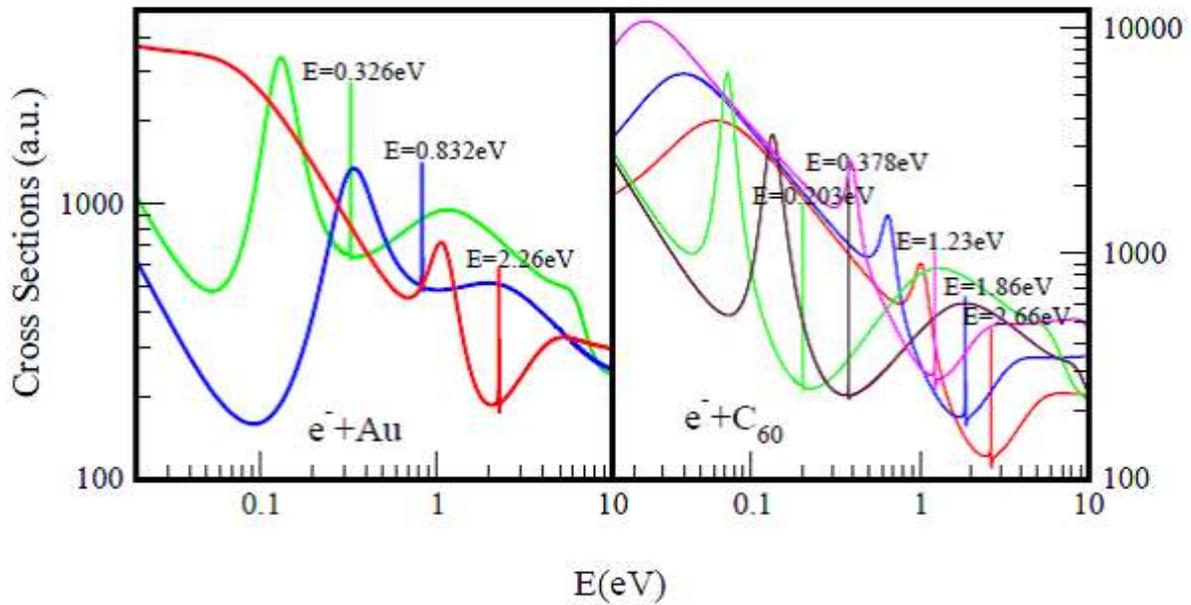

**Figure 3.1:** Total cross sections (a.u.) for electron elastic scattering from atomic Au (left panel) and the fullerene molecule $C_{60}$ (right panel) are contrasted. For atomic Au the red, blue and green curves represent TCSs for the ground, metastable and excited states, respectively. For the $C_{60}$ fullerene the red, blue and pink curves represent TCSs for the ground and the metastable states, respectively while the green and brown curves denote TCSs for the excited states. The very sharp resonances in both Figs. correspond to the $Au^-$ and $C_{60}^-$ anionic formation. The anionic BEs in the Figs. are intended to guide the eye; the complete values are also presented in Table 3.1 for better comparisons.

**Table 3.1:** Negative-ion binding energies (BEs) and ground states Ramsauer-Townsend (R-T) minima, all in eV extracted from TCSs of the atoms Au, Pt, At, Ti and Hf and the fullerene molecules $C_{60}$ and $C_{70}$. They are compared with the measured electron affinities (EAs) in eV. GRS, MS-$n$ and EXT-$n$ ($n$=1, 2) refer respectively to ground, metastable and excited states. Experimental EAs, EXPT and theoretical EAs, Theory are also included. The numbers in the square brackets are the references.

| System Z | BEs GRS | BEs MS-1 | BEs MS-2 | EAs EXPT | BEs EXT-1 | BEs EXT-2 | R-T GRS | BEs/EAs Theory |
|---|---|---|---|---|---|---|---|---|
| Au 79 | 2.26 | 0.832 | - | 2.309[4] 2.301[5] 2.306[6] | 0.326 | - | 2.24 | 2.262[25] |
| Pt 78 | 2.16 | 1.197 | - | 2.128[4] 2.125[7] 2.123[8] | 0.136 | - | 2.15 | 2.163[25] |
| At 85 | 2.41 | 0.918 | - | 2.416[9] | 0.115 | 0.292 | 2.43 | 2.38[15] 2.42[55] 2.51[56] 2..80[57] |
| $C_{60}$ | 2.66 | 1.86 | 1.23 | 2.684[10] 2.666[11] 2.689[12] | 0.203 | 0.378 | 2.67 | 2.663[17] 2.63[58] 2.57[59] |
| $C_{70}$ | 2.70 | 1.77 | 1.27 | 2.676[11] 2.72[13] 2.74[14] | 0.230 | 0.384 | 2.72 | 3.35[60] 2.83[60] |
| Ti 81 | 2.42 | - | - | 0.377[21] 0.075[22] | 0.066 | 0.281 | 2.40 | 0.27 [23] 0.291 [24] |
| Hf 72 | 1.68 | 0.525 | - | 0.178[26] | 0.017 | 0.113 | 1.67 | 0.114[27] 0.113[28] |

### 3.2 Ground State Fullerene Cross Sections

The main reason for following up with the fullerene molecules immediately after Section 3.1 is that excellent measured EAs from $C_{20}$ through $C_{92}$ are available in the literature. Benchmarked on the measured EAs of $C_{60}$ and $C_{70}$ as indicated under Section 3.1, the Regge pole method was used to calculate the ground state electron elastic TCSs of the fullerene molecules from $C_{20}$ through $C_{240}$[16,17]. It is noted here that in the paper [17] only the ground states anionic BEs were calculated and some of the fullerenes ground state BEs can be found there. The novelty and generality of the Regge pole approach is in the extraction of the anionic BEs from the calculated TCSs of the fullerenes; for ground state collisions these BEs yield the unambiguous and definitive challenging to calculate theoretically EAs. In [16, 17] the ground states anionic BEs of the fullerenes $C_{20}$ through $C_{92}$ were found to match in general excellently the measured EAs. Indeed, these results provided great credence to the ability of the Regge pole method to extract from the calculated TCSs reliable EAs of the fullerene molecules for the first time. The obtained agreement represented an unprecedented accomplishment by the Regge pole method, requiring no assistance whatsoever from either experiment or any other theory for the feat. This allowed the interpretation of the EAs of fullerenes as corresponding to the ground states anionic BEs calculated by the Regge pole method. For the fullerenes other theories continue to struggle to go beyond the theoretically simple $C_{20}$ and $C_{60}$ fullerenes.

The focus in this Section is on the ground state anionic BEs of the fullerenes from $C_{20}$ through $C_{240}$ [16, 17]. These ground states TCSs are typified by the red curve of the $C_{60}$ TCSs of Fig 3.1. Additional to the ground state curve the revealing metastable and excited states TCSs curves demonstrate the richness in structure of the fullerene TCSs (the larger fullerenes reveal more metastable and excited states TCSs than shown in the Fig. 3.1). For the $C_{20}$, $C_{24}$, $C_{26}$, $C_{28}$ and $C_{44}$ we have used the data of [16] to demonstrate the reliability of the Regge pole-calculated anionic BEs, since their measured EAs are available. The BE values in Table 3.2 cover a wider range of the fullerene anionic BEs than those shown in the papers [16, 17]. They include values where the experimental EAs are unavailable. In [16] the smaller fullerenes $C_{20}$, $C_{24}$, $C_{26}$, and $C_{28}$ as well as the larger fullerenes $C_{92}$ and $C_{112}$ were studied to assess the extent to which fullerenes behave like "big atoms" as suggested by Amusia [61]. These TCSs were found to be characterized generally by ground, metastable and excited states negative-ion formation, R-T minima and SRs. The ground states anionic BEs correspond to the measured EAs of the fullerenes, see Table 3.2 as well as Fig. 3.1 for the $C_{60}$ TCSs. The Regge pole method does not determine the orbital angular momentum of the attached electron. This is particularly important for the $C_{60}$ fullerene since there is an uncertainty in the literature regarding whether the $C_{60}$ EA of 2.66eV corresponds to an s-state or a p-state of the attached electron. Since the ground state anionic BE (EA) of $C_{60}$ is determined here, we believe that a structure-type calculation could use our calculated ground state anionic BE (EA) of $C_{60}$ to determine the ground-state fine-structure energies, since the metastable energies are also available. Thus the lingering question could be answered.

The Regge pole-calculated low-energy electron elastic TCSs for the ground and the first (highest) excited states of fullerenes are robust. For $C_{20}$ (smallest fullerene) the first excited state TCS (highest TCS) resembles that of atomic Au, see Fig. 3.1. Defining R as the value of the ratio of the second to the first R-T minima in the first excited state TCS of $C_{20}$, in [16] we explored the variation of R from $C_{20}$ through $C_{70}$. We found that for $C_{20}$ R (~1.4), greater than unity was close to that for Au (~2.) or Th (~1.9), indicative of atomic behavior, while for $C_{24}$ R was about 1.0. For $C_{70}$ R was less than 0.5 demonstrating strong departure from atomic behavior due to the significant polarization interaction in $C_{70}$; which also induces long-lived metastable anions in the $C_{70}$ TCSs. When probed with low-energy electrons, the results for $C_{20}$ exhibited fullerene behavior consistent with the view that fullerenes behave like "big atoms" [61]. The atomic behavior quickly disappears with the increase in the fullerene size. As seen from the Figures of [16], the behavior in $C_{28}$ is no longer atomic because R is less than unity. By $C_{92}$ the departure from atomic behavior has become significant, due to the increase in the polarization interaction in these larger systems.

For $C_{20}$ the excited state TCS [16] exhibits a deeper R-T minimum near threshold in comparison with the second R-T minimum, while the ground state TCS ends with a deep R-T minimum, wherein appears the dramatically sharp resonance representing the stable negative ion formed in the ground state during the collision, see also Fig. 3.1. These characteristic R-T minima, also observed in the Dirac R-matrix low-energy electron elastic scattering cross sections calculations for the heavy, alkali-metal atoms Rb, Cs and Fr [62], manifest that the important core-polarization interaction has been accounted for adequately in our calculation, consistent with the conclusion in [63]. The vital importance of the core-polarization interaction in low-energy electron collisions with atoms and molecules was recognized and demonstrated long ago, see for example [64] and references therein. In $C_{20}$ the TCSs are characterized by a ground, metastable and excited states TCSs. However, the $C_{24}$, $C_{26}$ and $C_{28}$ TCSs consist of more metastable and excited states TCSs. Suffice to state that the increased energy space determined mainly by the ground states BEs is conducive to the appearance of the polarization-induced metastable TCS in general. Indeed, these results reveal the complicated interplay between the R-T minima and the shape resonances.

Notably, in all the fullerene molecules investigated here, the ground states anionic BEs occur at the absolute R-T minima of the TCSs, see Fig. 3.1 for example. This facilitates considerably the determination of unambiguous and reliable EAs of the fullerene molecules. Also noted here is that generally the sharp resonances of the metastable TCSs lie between the ground states SRs and the dramatically sharp resonances of the ground states. In Table 3.2 we have presented various fullerene anionic BEs, but mainly ground states BEs and compared them with the measured EAs where these are available. The results demonstrate the power of the Regge pole method to calculate reliably the anionic BEs of fullerene molecules.

Clearly, the Regge pole approach, entirely new in the field of electron-cluster/fullerene collisions, implemented here represents a theoretical breakthrough in low-energy electron scattering investigations of fullerenes/clusters and complex heavy atoms. Its implementation should speed up the long overdue fundamental theoretical understanding of the mechanism underlying low-energy electron scattering from fullerenes, including heavy and complex atoms, leading to negative ion formation. These results should also help in the construction of the popular square-well potentials for the investigated fullerenes. Most important, its great strength is in the ability to produce reliable data without assistance from experiments and/or other theories.

**Table 3.2**: Fullerene ground (GR-S), metastable (MS-$n$, $n$ =1, 2, 3) and first excited (EXT-1), second excited (EXT-2) and third excited (EXT-3) anionic states binding energies (BEs). The measured EAs are represented as EXPT and the other theoretical values are denoted as Theory. All the energies are in eV and the numbers in the square brackets are the references. For most of the fullerenes the parameters "α" and "β" of the potential, Eq. (2) are tabulated in [17].

| System | EA EXPT. | BE(Ours) GR-S | BEs MS-1 | BEs MS-2 | BEs MS-3 | BEs EXT-1 | BEs EXT-2 | BEs EXT-3 | EA Theory |
|---|---|---|---|---|---|---|---|---|---|
| $C_{20}$ | 2.44[65] 2.60[66] 2.70[67] | 2.72 | 1.48 | - | - | 0.466 | - | - | - |
| $C_{24}$ | 3.750[66] 2.90 [67] | 3.79 | 2.29 | 1.41 | - | 0.428 | 0.801 | - | - |
| $C_{26}$ | 3.100[66] 2.95 [67] | 2.67 | 1.59 | - | - | 0.464 | - | - | - |
| $C_{28}$ | 2.80 [66] 3.00 [67] | 3.10 | 1.80 | - | - | 0.305 | 0.505 | - | - |
| $C_{44}$ | 3.30 [68] | 3.15 | 1.89 | 1.47 | - | 0.319 | 0.492 | - | - |
| $C_{74}$ | 3.28 [69] | 4.03 | 2.83 | 2.01 | 1.48 | 0.251 | 0.407 | 0.643 | |
| $C_{76}$ | 2.89 [69] | 2.79 | | | | | | | |
| $C_{78}$ | 3.10 [69] | 2.98 | | | | | | | |
| $C_{80}$ | 3.17 [69] | 3.28 | | | | | | | |
| $C_{82}$ | 3.14 [69] | 3.15 | | | | | | | 3.37[59] |
| $C_{84}$ | 3.05 [13] | 2.94 | | | | | | | |
| $C_{86}$ | ≥ 3.0[13] | 2.92 | | | | | | | |
| $C_{90}$ | ≥ 3.0[13] | 3.06 | | | | | | | |
| $C_{92}$ | ≥ 3.0[13] | 3.09 | 2.35 | 1.58 | - | 0.266 | - | - | |
| $C_{100}$ | - | 3.67 | 2.70 | 2.04 | - | 0.242 | 0.379 | 0.531 | - |
| $C_{112}$ | - | 3.31 | 2.53 | 1.73 | - | 0.243 | 0. 315 | 0.519 | |
| $C_{120}$ | - | 3.74 | 2.97 | 2.04 | 1.58 | 0.244 | 0.372 | 0.576 | |
| $C_{124}$ | - | 3.06 | 2.30 | 1.71 | - | 0.289 | 0.393 | 0.569 | |
| $C_{132}$ | - | 3.59 | 2.60 | 1.93 | - | 0.251 | 0.338 | - | |
| $C_{136}$ | - | 3.75 | 2.64 | 2.19 | 1.67 | 0.260 | 0.345 | 0.488 | |
| $C_{140}$ | - | 3.94 | 3.06 | 2.23 | 1.75 | 0.360 | 0.562 | 0.716 | |
| $C_{180}$ | - | 3.75 | 2.64 | 2.19 | 1.67 | 0.260 | 0.345 | 0.488 | 2.61[70] |
| $C_{240}$ | - | 4.18 | - | - | - | - | | | 3.81[71] 2.32[72] |

## 3.3 Cross sections for the large atoms Hf, Pt, Au, Ti and At

In the context of the viewpoints 1) and 2) of the Introduction, it is appropriate to discuss the measured EAs of the large atoms Hf(72), Pt(78), Au(79), Ti(81) and the radioactive At (85) in an attempt to understand the meaning of the measured EAs of the Hf and Ti atoms (the numbers within the brackets are the Z's). That is, do their EAs correspond to electron BEs in the ground, the metastable or the excited states of the formed anions during the collision? As seen from Table 3.1 the meaning of the EAs of Au, Pt and At is clear, namely it corresponds to the ground state anionic BEs of the formed anions during the collisions. However, for Hf and Ti the meaning lacks definitiveness.

For clarity the Figure 3.3 shows the TCSs for the Hf atom; a similar Figure was obtained for the Ti atom. As seen from the Figure, it is difficult to understand any selection of the anionic BEs other than the ground state anionic BE as the EA of Hf. A similar argument applies to the Ti atom. The measured EA of Hf at 0.178eV[26], the Regge pole-calculated SR of 0.232eV, the RCI EA of 0.114eV[27] and the Regge pole-calculated second excited state anionic BE of 0.113eV[28] are reasonably close together. The highest excited state BE of Hf is at 0.017eV[29]. The TCSs for Hf presented in Fig. 3.3 demonstrate the additional presence to the above discussed anionic BEs, a metastable TCS (green curve) and a ground state TCS (pink curve) with anionic BEs of 0.525eV and 1.68eV, respectively. Indeed, here we are faced with the problem of interpretation of what is meant by the EA.

As indicated in the Introduction, for the Ti atom two measurements obtained its EAs as 0.377eV[21] and 0.075eV[22]. The former value is close to various theoretical calculations [23, 24], including the Regge pole-calculated BE of the second excited anionic state, namely 0.281eV [56], see comparisons in Table 3.1. However, the measured value of 0.075eV[22] and the Regge pole-calculated BE of the highest excited state of the formed Ti⁻ anion, 0.0664eV are close together. We note that the Regge pole-calculated ground state anionic BE of Ti is 2.42eV [56], very close to that of the At atom. Clearly, the results of the Hf and Ti atoms are difficult to interpret without a rigorous theoretical data as discussed under Section 3.4 dealing with the lanthanide atoms.

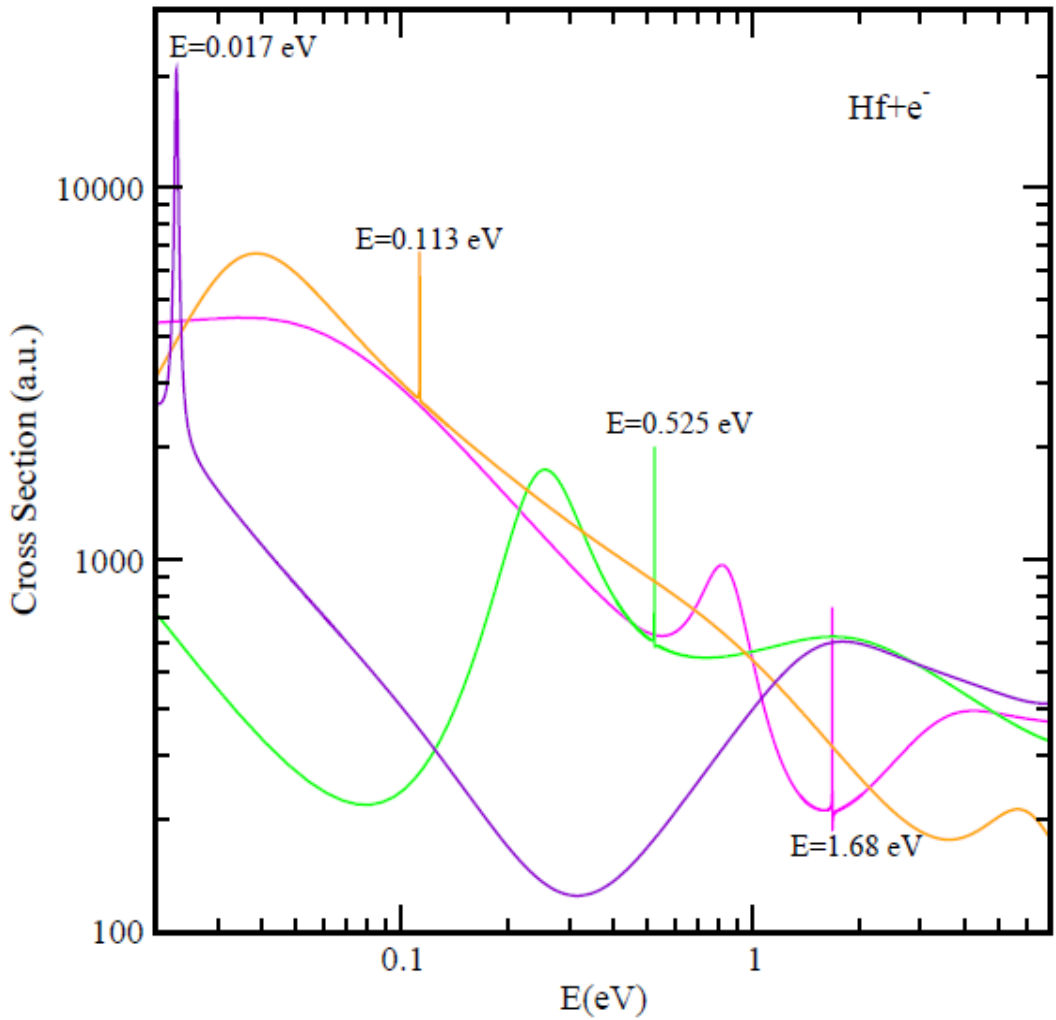

**Figure 3.3:** Total cross sections (a.u.) for electron elastic scattering from Hf. The pink, green, orange and purple curves represent the TCSs for the ground, metastable and the two excited states, respectively. The dramatically sharp resonances correspond to the Hf⁻ anionic formation during the collisions.

### 3.4 Cross sections for the lanthanide atoms

The EA provides a stringent test of theory when the theoretical EAs are compared with those from reliable measurements. This statement holds strongly in the case of the lanthanide atoms. The general problem of interpretation of the measured EAs of the lanthanide atoms has been exposed in [29] as well as elucidated through the rigorous Regge pole method [73]. Appropriately, we begin this section by placing in perspective the existing measurements/calculations of the EAs of the lanthanide atoms. Low-energy electron elastic collision cross sections for the lanthanide atoms, La through Lu were first investigated using the CAM (Regge pole) method [33]. Unfortunately, the investigation was limited to the near threshold energy region, $0.0 \leq E \leq 1.0$ eV and focused upon the comparison with the existing measured and theoretical EAs and never questioned the meaning of the EAs. The CAM calculated TCSs were found to be characterized generally by dramatically sharp resonances whose energy positions were identified with the measured/calculated EAs of the lanthanide atoms. The extracted EAs from the TCSs varied from a low value of 0.016eV for Tm to a high value of 0.631eV for atomic Pr. In that paper [33] one sees the effective use of the Regge Trajectories and the Im L (L is the complex angular momentum) in analyzing and interpreting the results. Also, the lanthanide parameters "α" and "β" of the potential, Eq. (2) are tabulated in that paper [33].

Subsequently, when the energy range was increased from 1.0eV to 10.0eV, ground, metastable and excited states anionic BEs were clearly revealed and delineated. Then, the question persisted: do the measured EAs of the lanthanide atoms correspond to the anionic BEs of electron attachment in the ground, metastable or excited states of the formed anions during the collision? For atomic Eu the resonance at E = 0.116eV with Im L = $7.6 \times 10^{-6}$ of the Fig. 7 of [33] should be compared with the results of Fig. 3 of [30] reproduced here for convenience as Fig. 3.4(left panel). In Fig. 13 depicting the TCSs of Tm [33] the dramatically sharp resonance at E = 0.016eV with Im L = $3.4 \times 10^{-5}$ should be viewed in the context of the recent Fig. 3 of [30], also presented here as Fig. 3.4(right panel).

The lanthanide and the Hf atoms provide clear cases of the ambiguous and confusing measured and/or calculated EA values. As examples, for Eu we focus on the ground state, pink curve with the BE value of 2.63 eV and the blue curve with the BE of 0.116 eV, corresponding to an excited state TCS. The measured EA (0.116 eV)[3] is in outstanding agreement with the excited state BE value above and the RCI calculated EA (0.117 eV)[36], see Table 1 of [30]. The metastable BE value of 1.08 eV, red curve in Fig, 3.4(left panel) agrees excellently with the measured

EA (1.053 eV)[35]. This clearly demonstrates the ambiguous and confusing meaning of the measured EA of Eu by Refs. [3] and [35]. Does the EA of Eu correspond to the BE of electron attachment in the metastable state or in the excited state of the formed anion during the collision? Similarly with the case of the Tm atom; the Regge pole calculated ground and excited states BEs are respectively 3.36 eV and 0.274 eV. The measured EA of Tm is 1.029 eV[37] and agrees excellently with the Regge pole calculated metastable state BE value of 1.02 eV, green curve in the Fig. 3.4(right panel). Accordingly, here the meaning of the measured EA of Tm corresponds to the BE of the metastable state. In both Eu and Tm atoms the meaning of the measured EAs is ambiguous and confusing as well.

A comment is appropriate here regarding the importance or unimportance of Relativistic effects in the calculation of the EAs using the Regge pole method. With a relatively high Z of 63, but small measured EA of 0.116 eV [3], the Eu atom provides a stringent test of the nonrelativistic CAM method when its prediction (EA=0.116eV) [33] is compared with that calculated using the MCDF-RCI (EA=0.117 eV) [36]. The interpretation aside, the results demonstrate the unimportance of relativistic effects in the calculation of the EA of Eu. Indeed, the EAs calculated using structure-based theoretical methods tend to be riddled with uncertainty and lack definitiveness for complex multi-electron systems and fullerene molecules. For instance, Relativistic effects in gold chemistry were investigated by Wesendrup et. al [74] who performed large-scale fully Relativistic Dirac–Hartree–Fock and MP2 calculations as well as Nonrelativistic pseudopotential calculations and obtained the EAs of 2.19 eV and 1.17 eV, respectively. These values should be contrasted with the CAM calculated value of 2.263 eV [25] and compared with the measured EAs of Au in Table 3.1. Also of importance is the review on relativistic effects in homogeneous gold catalysis [75].

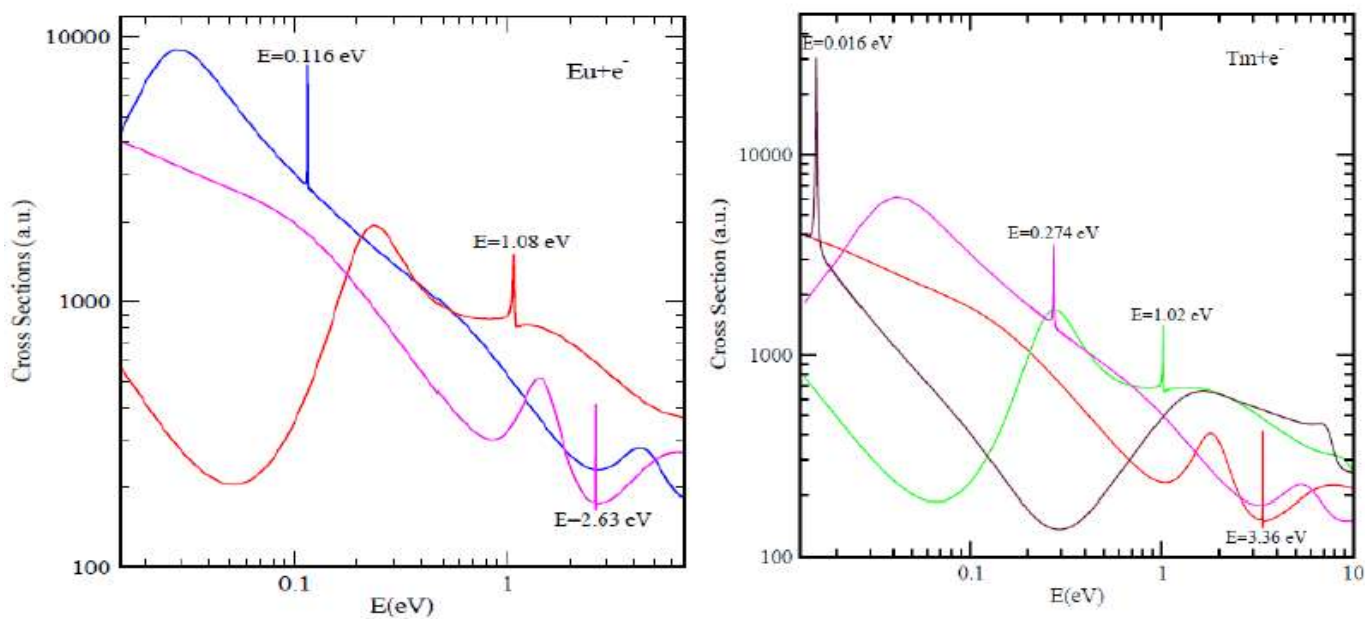

**Figure 3.4:** Total cross sections (a.u.) for electron elastic scattering from atomic Eu (left panel) and Tm (right panel). For Eu the pink, red and blue curves represent the TCSs for the ground, metastable and the excited states, respectively. For Tm atom the red, green, pink and black curves represent the TCSs for the ground, metastable and the two excited states, respectively. The dramatically sharp resonances in both figures correspond to the Eu⁻ and Tm⁻ negative-ions formed during the collision.

## 3.5  Cross Sections for the Actinide atoms

In [54] we investigated the low-energy electron scattering from the radioactive actinide atoms Th, Pa, U, Np and Pu through the elastic TCSs calculations. The objective was to delineate and identify the characteristic resonance structures as well as to understand and assess the reliability of the existing theoretical EAs. The recent measurement of the EA value of Th warrants some remark. There is no reason whatsoever for the selective comparison of data by the experiment; there are calculated EAs in the literature [36,76]. Particularly interesting in the study above[54] is the finding for the first time that the TCSs for atomic Pu exhibited fullerene molecular behavior near threshold through the TCS of the highest excited state, while maintaining the atomic character through the ground state TCS. Also, the first appearance of the near threshold deep R-T minimum in the actinide TCSs was first identified in the TCSs of atomic Pu, see Fig. 6 of [54].

Figure 3.5, taken from Ref. [54] with a slight modification due to recalculation presents the TCSs for atomic Th (top figure) and U (bottom figure). They typify the TCSs of the complex multi-electron actinide atoms. Importantly, they are characterized by dramatically sharp resonances representing negative-ion formation in the ground, metastable and excited anionic states, R-T minima and SRs. In both Figures the red curves represent electron attachment in the ground states while the pink curves denote the highest excited states. For Th, Fig. 3.5 (top) the measured and the calculated EA values are 0.608 eV and 0.599 eV[18], respectively. These values are close to the Regge pole-calculated

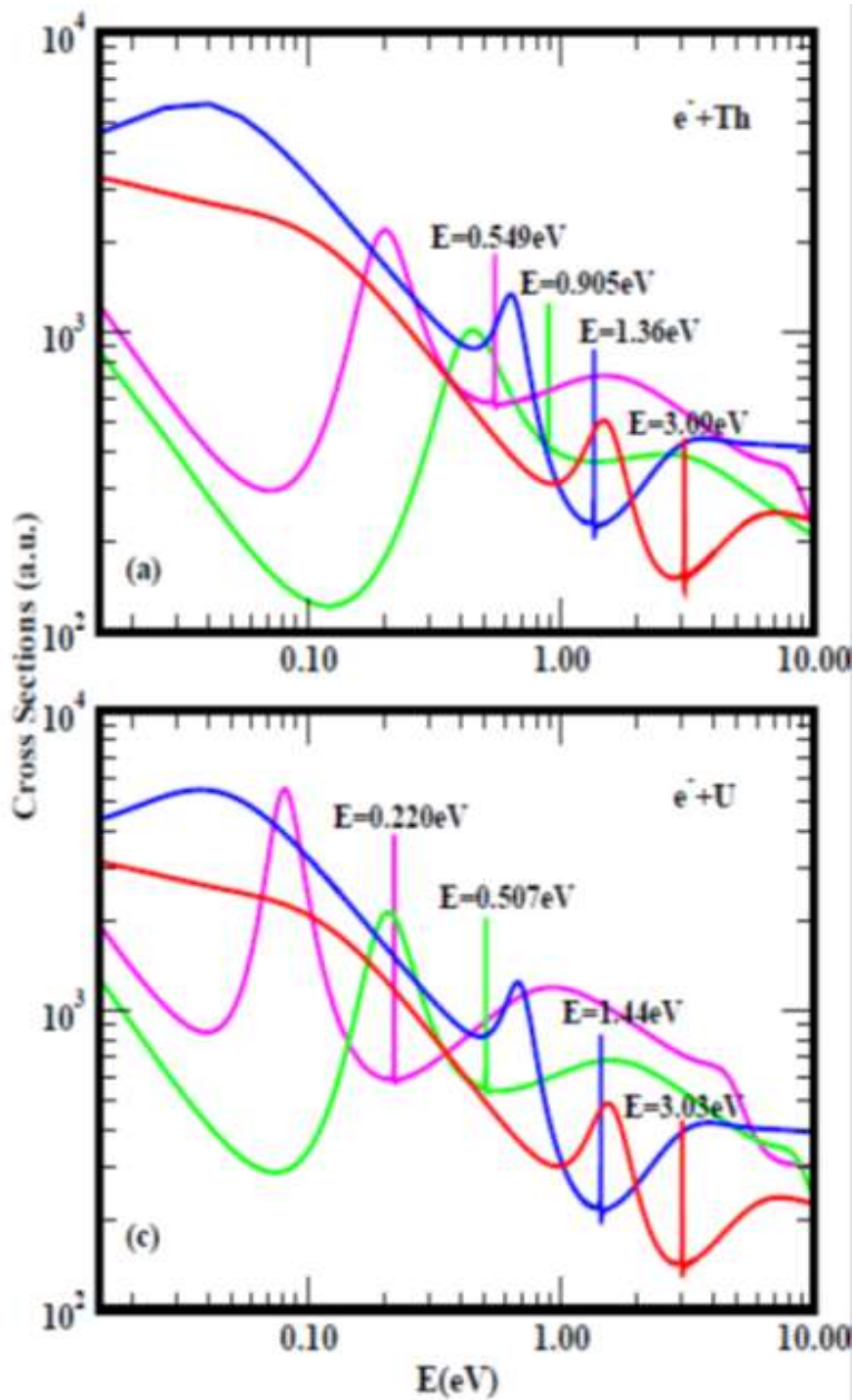

**Figure 3.5:** Total cross sections (TCSs) for atomic Th (top) and U(bottom) Figures. In this paper the relevant curves in both TCSs are the ground states (red curves) and the excited states (pink curves). The dramatically sharp resonances in both Figs. with attendant BEs represent electron attachment. These BE values are intended to guide the eye; the complete values are presented in Table 3.5. There are also shape resonances next to these sharp peaks

anionic BE of the second excited state, pink curve (0.549 eV). Close to this value there is a SR at 0.61 eV defined by the blue curve; the ground state anionic BE is at 3.09 eV. Not shown is the highest excited state curve with anionic BE value of 0.149 eV. Importantly, here we note the clear atomic behavior exhibited by the Th TCSs (pink curve) as expected [16]. However, the pink curve in the U TCSs shows strong fullerene behavior [16]. The EAs of U have been measured very recently to be 0.315 eV [19] and 0.309 eV [20] as well as calculated to be 0.232 eV [20]. These values are close to the Regge pole anionic BE value of 0.220 eV for the highest excited state, see Table 3.5 and Ref. [54] for additional comparisons. Here we do not understand the inconsistency in the meaning of the EAs in Figs. 3.1 and 3.5, namely as corresponding to the BEs of electron attachment in the ground and the excited anionic states, respectively

Of particular interest and importance here are the contrasted TCSs for atomic Am and Lr in [77]. In that paper, the polarization-induced TCS (brown curve) of Am exhibited a deep R-T minimum near threshold. This brown curve behaves similarly to that in the TCSs of Pu, while in Lr it has already flipped over to a SR. Where does the actual flipping take place? This is the subject of the Section 3.7. To understand the measurement [38] we need the previously unavailable data for the Cf, Fm and Md atoms to determine where the actual flipping takes place. There are no measured EAs for the actinide atoms beyond U to compare our BEs with. However, theoretical EAs are available [36, 76, 78, 79] and these have been compared with our data for the actinide atoms [54, 77, 80], see also Table 3.5. For atomic Lr the EA values of 0.310 eV[78] and 0.295eV[36] are very close to the Regge pole BE of the highest excited state, namely 0.321eV. These values should guide the reliable measurement of the EA of Lr. We have brought together the data for most of the actinide atoms in Table 3.5 mainly for convenience of analysis. Hitherto fore, they were scattered all over the literature. Importantly, the recent data from [80] are crucial for understanding Section 3.7.

**Table 3.5:** Negative ion binding energies (BEs) in eV and energy positions of ground-state Ramsauer–Townsend (R-T) minima, in eV obtained from the TCSs for the actinide atoms from Pu through Lr. Additionally, included for comparison are the data for Au, Th and U. GRS, MS-$n$ and EXT-$n$ ($n$ = 1, 2) represent ground, metastable and excited states, respectively. The experimental EAs, EXPT and the theoretical EAs, including RCI [36] and GW [76] are also presented. The numbers in the square brackets are the references.

| System/ Z | BEs GRS | BEs MS-1 | BEs MS-2 | EAs EXPT | BEs EXT-1 | BEs EXT-2 | R-T GRS | BEs/EAs Theory | EAs [36] | EAs [76] |
|---|---|---|---|---|---|---|---|---|---|---|
| Au 79 | 2.26 | 0.832 | - | 2.309 [4] | 0.326 | - | 2.24 | 2.262 [25] | - | - |
| Th 90 | 3.09 | 1.36 | 0.905 | 0.608 [18] | 0.149 | 0.549 | 3.10 | 0.599 [18] | 0.368 | 1.17 |
| U 92 | 3.03 | 1.44 | - | 0.315[19] 0.309[20] | 0.220 | 0.507 | 3.01 | 0.232[20] 0.175[36] | 0.373 | 0.53 |
| Pu 94 | 3.25 | 1.57 | 1.22 | N/A | 0.225 | 0.527 | 3.26 | - | 0.085 | −0.503 −0.276 |
| Am 95 | 3.25 | 1.58 | 0.968 | N/A | 0.243 | 0.619 | 3.27 | - | 0.076 | 0.103 0.142 |
| Bk 97 | 3.55 | 1.73 | 0.997 | N/A | 0.267 | 0.505 | 3.53 | - | 0.031 | −0.503 −0.276 |
| Cf 98 | 3.32 | 1.70 | 0.955 | N/A | 0.272 | 0.577 | 3.34 | - | 0.018 0.010 | -0.777 -1.013 |
| Es 99 | 3.42 | 1.66 | 0.948 | N/A | 0.272 | 0.642 | 3.44 | - | 0.002 | 0.103 0.142 |
| Fm 100 | 3.47 | 1.79 | 1.02 | N/A | 0.268 | 0.623 | 3.49 | - | - | 0.597 0.354 |
| Md 101 | 3.77 | 1.81 | 0.996 | N/A | 0.259 | 0.700 | 3.78 | | - | 1.224 0.978 |
| No 102 | 3.83 | 1.92 | 1.03 | N/A | 0.292 | 0.705 | 3.85 | - | - | −2.302 −2.325 |
| Lr 103 | 3.88 | 1.92 | 1.10 | N/A | 0.321 | 0.649 | 3.90 | 0.310 [78] 0.160 [78] 0.476 [79] | 0.465 0.295 | −0.212 −0.313 |

It is now clear why many existing experimental measurements and sophisticated theoretical calculations have considered the anionic BEs of the stable metastable and/or excited negative ion formation to correspond to the EAs of the considered lanthanide and actinide atoms. This is contrary to the usual meaning of the EAs found in the standard measurement of the EAs of such complex systems as atomic Au, Pt and the radioactive At as well as of the fullerene molecules. In these systems, the EAs correspond to the ground state BEs of the formed negative ions. The negative ions obtained here are also important in catalysis.

### 3.6 Fullerene Negative-Ion Catalysis
#### 3.61 Overview

The extensive and crucial applications of fullerenes in science, nanotechnology and industrial research as well as in astrophysics have motivated this study. The acceptor material used particularly in modern organic solar cells is usually a fullerene derivative [81]. Understanding the stability and degradation mechanism of organic solar cells is essential before their commercialization. Toward this end designing polymers and fullerenes with larger electron affinity (EA) has been proposed [82]. This motivated our first ever study of the large fullerenes [83] as well as the present study to search for fullerenes with larger EAs. The rich long-lived metastable resonances that characterize the Regge pole-calculated large fullerenes TCSs presented for the first time in [83] support the important conclusion that the experimentally detected fullerene isomers correspond to the metastable states [84] and further confirm the need to identify and delineate the resonance structures in gentle electron scattering.

The fundamental mechanism underlying atomic negative-ion catalysis was proposed by our group in the context of muon catalyzed nuclear fusion [85, 86]. The mechanism involves anionic molecular complex formation in the transition state (TS), with the atomic negative ion breaking the hydrogen bond strength. The mechanism has been demonstrated in the synthesis of $H_2O_2$ from $H_2O$ catalyzed using the Au⁻ and Pd⁻ anions to understand the experiments of Hutchings and collaborators [87, 88, 89], in the catalysis of light, intermediate and heavy water to the corresponding peroxides [90] and in the oxidation of methane to methanol without the $CO_2$ emission [91]. More recently, the experiment [89] has used the less expensive atomic Sn for possible water purification in the developing world. In this context we explored [92] the effectiveness of the fullerene anions $C_{20}^-$ to $C_{136}^-$ in the catalysis of water oxidation to peroxide and water synthesis from $H_2$ and $O_2$ hoping to find inexpensive effective negative-ion catalysts.

### 3.62. Results

The electron elastic TCSs for the typical large fullerenes $C_{100}$, $C_{120}$ and $C_{140}$ demonstrate negative-ion formation [92, 93, 94] with significant differences among their EAs, namely 3.67eV, 3.74eV and 3.94eV, respectively, see also Table 3.2. It is now clear that the ground state anionic BEs located at the absolute R-T minima of the ground state TCSs yield the challenging to calculate theoretically EAs. Indeed, the R-T minimum provides an excellent environment that is conducive to negative-ion catalysis and the creation of new molecules. The underlying physics in the fullerene TCSs has already been explained in [93, 94]. The obtained results are consistent with the observation that low-energy electron-fullerene interactions are characterized by rich resonance structures [95, 96] and that the experimentally detected fullerene isomers correspond to the metastable TCSs [84]. They also support the conclusion that the EAs of fullerene molecules are relatively large. The results of [83] including those presented in Table 3.2 should satisfy part of the requirement to increase fullerene acceptor resistance to degradation by the photo-oxidation mechanism through the use in organic solar cells of fullerenes with high EAs [82]. The extracted EAs from the TCSs could also be used to construct the widely used simple model potentials for the fullerene shells, including endohedral fullerenes [97 -105] as well as in the study of the stability of An@$C_{40}$(An = Th, -----,Md) [106]. Notably, the EAs are at the hearts of many of the model potentials. Indeed, the rich resonance structures in the fullerenes TCSs and their large EAs explain the tendency of fullerenes to form compounds with electron-donor anions and their vast applications as well.

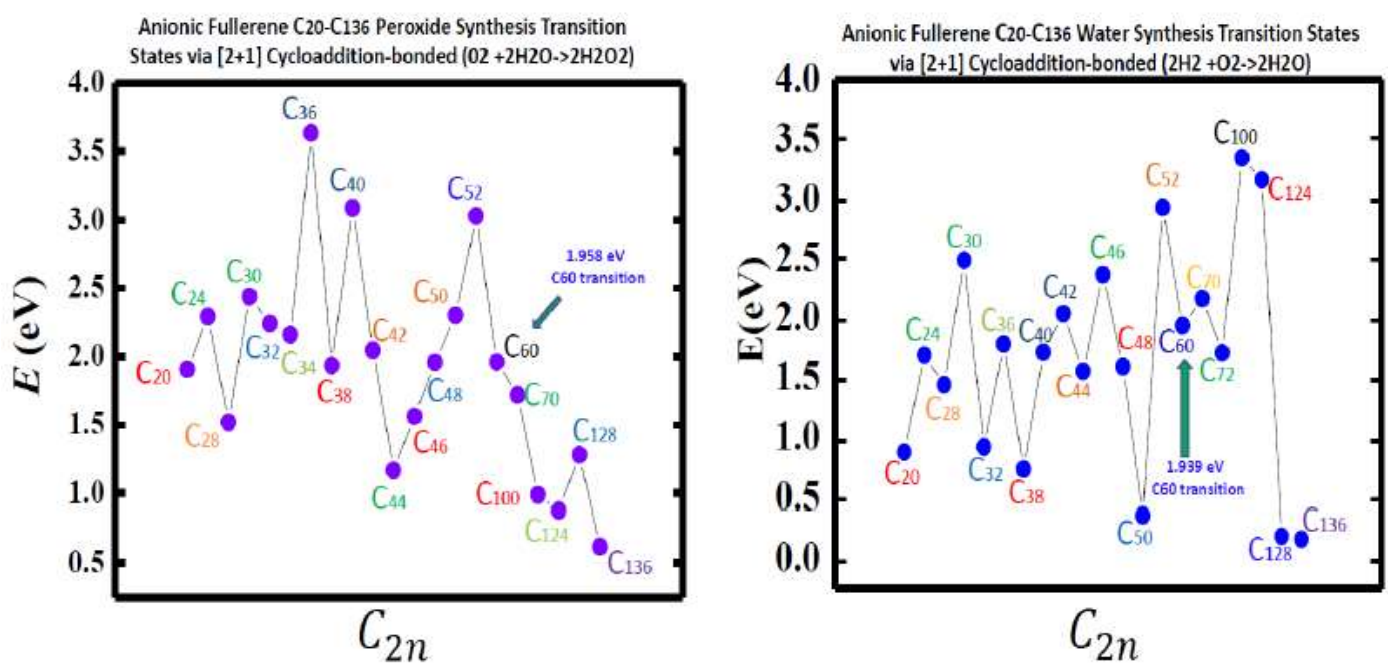

**Figure 3.6:** Transition state energy barriers of anionic fullerenes sizes $C_{20}^-$ to $C_{136}^-$ for catalyzing water oxidation to peroxide (left panel) and catalyzing hydrogen and oxygen synthesis to water (right panel).

The utility of the generated fullerene anions has been demonstrated in the catalysis of water oxidation to peroxide and water synthesis from $H_2$ and $O_2$ using the anionic fullerene catalysts $C_{20}^-$ to $C_{136}^-$ [92]. Figures 3.6 taken from [92] demonstrate the Density Functional Theory (DFT) calculated TS energy barriers for both processes. DFT and dispersion corrected DFT approaches have been employed for the TS evaluations. Geometry optimization of the structural molecular conformation utilized the gradient-corrected Perdew-Burke-Ernzerhof parameterizations [107] of exchange-correlation as implemented in DMol3 [108]. DFT calculated energy barriers reduction in the oxidation of $H_2O$ to $H_2O_2$ catalyzed using the anionic fullerene catalysts $C_{20}^-$ to $C_{136}^-$ are shown in the Fig. 3.6 (left panel). The results in Fig. 3.6 (right panel), also from [92] are for the water synthesis from $H_2$ and $O_2$ catalyzed using the anionic fullerene catalysts $C_{20}^-$ to $C_{136}^-$ as well. For both water oxidation and water synthesis DFT TS calculations found the $C_{52}^-$ and $C_{60}^-$ anions to be numerically stable and the $C_{36}^-$ and $C_{100}^-$ anions to increase the energy barriers the most in the water oxidation to $H_2O_2$ and water synthesis using $H_2$ and $O_2$, respectively. The $C_{136}^-$ anion has proved to be the most effective in reducing the energy barrier significantly when catalyzing both water oxidation to peroxide and synthesis from $H_2$ and $O_2$. Importantly, a single large fullerene such as the $C_{136}$, or even the $C_{74}$ could replace the Au, Pd and Sn atoms in the catalysis of $H_2O_2$ from $H_2O$ in the experiments [87-89] acting as a multiple-functionalized catalyst. These fullerenes have their metastable BEs close to the EAs of the atoms used in the experiments. Thus an inexpensive dynamic water purification system for the developing world could be realized [89].

Indeed, the utility of the fullerene molecular anions has been demonstrated in the catalysis of water oxidation to peroxide and water synthesis from $H_2$ and $O_2$ using the catalysts $C_{20}^-$ to $C_{136}^-$. DFT TS calculations found $C_{52}^-$ and $C_{60}^-$ anions numerically stable for both. The $C_{136}^-$ anion has proved to be the most effective in reducing the energy barrier significantly when catalyzing both water oxidation to peroxide and synthesis.

### 3.7 Atomic Structure and Dynamics of Bk and Cf: Experiment versus Theory

The recent experiment [38] using nanogram material identified a weak spin-orbit-coupling in atomic Bk while a jj coupling scheme described atomic Cf. It concluded that these observations strengthen Cf as a transitional element in the actinide series. Here the Regge pole-calculated low-energy electron elastic TCSs for Bk and Cf atoms are used as novel validation of the experimental observation through the sensitivity of R-T minima and SRs to the electronic structure of these atoms.

In Figures 3.7 we present the Regge pole-calculated electron elastic TCSs for the Bk (left panel) and Cf (right panel) actinide atoms. As seen the TCSs are characterized generally by dramatically sharp resonances, representing ground, metastable and excited states negative-ion formation, SRs (broad peaks) and R-T minima. Also, the highest excited states TCSs (green curves) exhibit fullerene molecular behavior [16]. The energy positions of the sharp resonances, well delineated correspond to the BEs of the formed negative ions during the electron collisions with the Bk and Cf atoms. Each figure consists of BEs of the ground (red curve), metastable (blue and orange curves) and excited (green and brown curves) states TCSs. At first glance these TCSs appear a little complicated. However, they can be readily understood if each curve is discussed separately, see also [93, 94] for example. The R-T minima manifest the effects of the polarization interaction [62], while the SRs convey the trapping effect of the centrifugal potential. The underlying physics has already been discussed in [93, 94]; it will not be repeated here.

For our objective here we focus mainly on the polarization-induced TCSs (orange curves) and the ground state TCSs in both Figs. 3.7. For a better understanding and appreciation of the results, it is appropriate to place in perspective the polarization-induced TCSs that are characterized by a deep R-T minimum near threshold in the TCSs of Bk and a pronounced SR very close to threshold in the Cf TCSs. The polarization-induced TCS with the deep R-T minimum near threshold first appeared in the actinides TCSs through the atomic Pu TCSs [54]. It was attributed to the size effect and the first 6d-orbital collapse impacting the polarization interaction significantly. The first 6d-collapse occurred in the transition Np[Rn]$7s^25f^46d$ to Pu[Rn]$7s^25f^6$. This caused the ground state anionic BE of the Np atom to increase from 3.06eV to 3.25eV in Pu. Also the anionic BE of the first metastable state increased from 1.47eV in Np to 1.57eV in Pu, see also Table 1 of Ref. [54]. It is the increase in the ground state energy space that facilitated the first appearance of the polarization-induced metastable TCS with the deep R-T minimum near threshold to appear in the Pu TCSs. This R-T minimum in the Pu TCSs continued through the Bk TCSs [77]. Indeed, the increase in the number of polarization-induced TCSs with size has already been clearly demonstrated in the fullerene molecular TCSs [83, 92], see also Fig. 1 of [54].

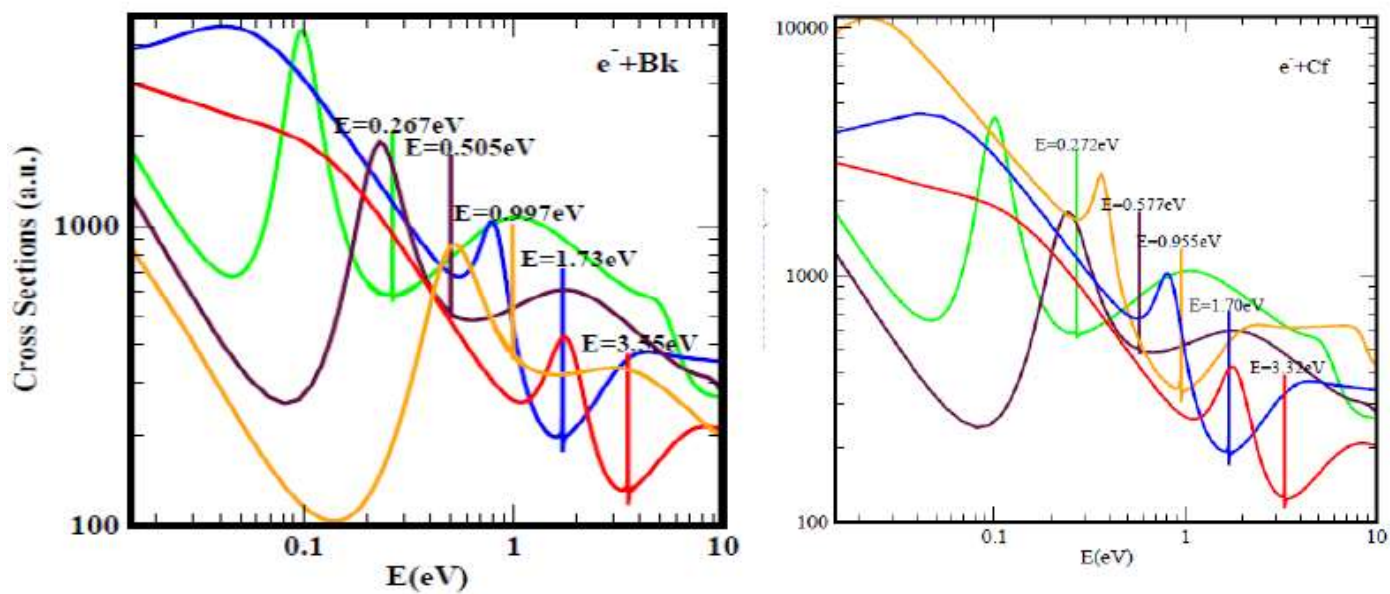

Figure 3.7: Total cross sections (a.u.) for electron elastic scattering from atomic Bk (left panel) and Cf (right panel). For both Bk and Cf the red and the blue and orange curves represent TCSs for the ground state and the metastable states, respectively. The green and the brown curves denote TCSs for the excited states. The orange curve with a deep R-T minimum in the Bk TCSs is the polarization-induced TCS due to size. In the Cf TCSs the orange curve has flipped over to a pronounced shape resonance very close to threshold. The dramatically sharp resonances in both figures correspond to the Bk⁻ and Cf⁻ anions formed during the collisions. The labeled BEs are intended to guide the eye; the complete values are presented in Table 3.5 for clarity.

The second 6d-orbital collapse occurs in the transition Cm[Rn]$7s^25f^76d$ to Bk[Rn]$7s^25f^9$. To facilitate the discussion we have included for convenience Table 3.7; the data have been taken from [54, 77, 80]. Table 3.7 shows that the ground state anionic BE increased significantly from 3.32eV in Cm to 3.55eV in Bk, thereby widening the energy space for the flip over to take place. Subsequently, the ground state anionic BE dropped from 3.55eV in Bk to 3.32eV in Cf. Similarly, for the first metastable states the anionic BEs increased from 1.57eV in Cm to 1.73eV in Bk, while it decreased to 1.70eV in Cf. In Table 3.7 it is informative to look at mainly the ground state energy values. Briefly, in Bk the ground state anionic BE is 3.55 eV; it decreased to 3.32 eV in Cf after the R-T minimum flipped over to a SR very close to threshold. This is indicative of the smaller energy space required by the SR compared to the R-T minimum. On the other hand, in Pu the ground state anionic BE is 3.25 eV having increased from 3.06 eV in Np to accommodate the first appearance of the polarization-induced TCS with the deep R-T minimum in Pu. Here we also see the appearance of the 1.22 eV BE as MS-2; the Figures are also quite informative.

**Table 3.7:** Negative-ion binding energies (BEs), in eV extracted from the TCSs for the actinide atoms Np through Fm are taken from [54, 77, 80]. GRS, MS-*n* and EX-*n* (*n*=1, 2) represent respectively ground, metastable and excited anionic states. The experimental EAs, EXPT, denoted by N/A are unavailable.

| Z | Atom | BEs GRS | EAs EXPT | BEs MS-1 | BEs MS-2 | BEs EX-2 | BEs EX-1 |
|---|------|---------|----------|----------|----------|----------|----------|
| 93 | Np | 3.06 | N/A | 1.47 | - | 0.521 | 0.248 |
| 94 | Pu | 3.25 | N/A | 1.57 | 1.22 | 0.527 | 0.225 |
| 95 | Am | 3.25 | N/A | 1.58 | 0.968 | 0.619 | 0.243 |
| 96 | Cm | 3.32 | N/A | 1.57 | 1.10 | 0.519 | 0.258 |
| 97 | Bk | 3.55 | N/A | 1.73 | 0.997 | 0.505 | 0.267 |
| 98 | Cf | 3.32 | N/A | 1.70 | 0.955 | 0.577 | 0.272 |
| 99 | Es | 3.42 | N/A | 1.66 | 0.948 | 0.642 | 0.272 |
| 100 | Fm | 3.47 | N/A | 1.79 | 1.02 | 0.623 | 0.268 |

The flip over of the near threshold Ramsauer-Townsend minimum from the Bk polarization-induced metastable TCS to a pronounced shape resonance very close to threshold in the Cf metastable TCS provides a sensitive probe of the electronic structure and dynamics of these atoms, thereby permitting the first ever use of the R-T minimum and the SR as novel confirmation of Cf as a transitional element in the actinide series, consistent with the experimental observation [38]. Indeed, the rigorous Regge pole method requires no assistance whatsoever from either experiments or any other theory for the remarkable feat, namely of probing reliably the electronic structure of these complicated actinide atoms.

## 4.0  Summary and Conclusions

The Regge pole-calculated low-energy electron elastic total cross sections (TCSs) of complex heavy multi-electron systems are characterized generally by dramatically sharp resonances manifesting negative-ion formation. These yield directly the anionic binding energies (BEs), the shape resonances (SRs) and the Ramsauer-Townsend(R-T) minima. From the TCSs unambiguous and reliable ground, metastable and excited states negative-ion BEs of the formed anions during the collisions are extracted and compared with the measured and/or calculated electron affinities (EAs) of the atoms and fullerene molecules. The novelty and generality of the Regge pole approach is in the extraction of rigorous negative-ion BEs from the TCSs, without any assistance whatsoever from either experiment or any other theory. The EA provides a stringent test of theoretical calculations when their results are compared with those from reliable measurements. For ground states collisions the Regge pole-calculated negative ion BEs correspond to the challenging to calculate theoretically EAs, yielding outstanding agreement with the standard measured EAs for Au, Pt and the highly radioactive At atoms as well as for the $C_{60}$ and $C_{70}$ fullerenes. For $C_{20}$ through $C_{92}$ fullerenes our Regge pole-calculated ground-state anionic BEs matched in general excellently the measured EAs. These results give great credence to the power and ability of the Regge pole method to produce unambiguous and reliable ground state anionic BEs of complex heavy systems through the TCSs calculation.

The meaning of the measured EAs of multi-electron atoms and fullerene molecules has also been considered here within the context of two prevailing viewpoints:
1) The first considers the EA to correspond to the electron BE in the ground state of the formed negative ion during collision; it is exemplified by the measured EAs of Au, Pt and At atoms and the fullerene molecules from $C_{20}$ through $C_{92;}$ and

2) The second view identifies the measured EA with the BE of electron attachment in an excited state of the formed anion. The measured EAs of Ti, Hf, lanthanide and actinide atoms provide representative examples of this viewpoint.

This experimental breakthrough [38], including the recent first ever measurements of the EAs of the highly radioactive element At[9], as well as the Th [18] and U[19, 20] atoms represent significant advances in the measurements of the challenging highly radioactive elements. And more such measurements of other radioactive atoms can be expected in the near future. Consequently, reliable theoretical predictions are essential for a fundamental understanding of the underlying physics. Here we have presented an entirely new approach to the validation of the experimental observation in [38], namely through the behavior of the R-T minima and the SRs in the metastable electron elastic TCSs of atomic Bk and Cf. Finally, with the available ground, metastable and excited negative-ion BEs calculated here for the multi-electron atoms and the fullerene molecules, sophisticated theoretical methods such as the Dirac R-matrix, Coupled-Cluster method, MCDHF, MCDF-RCI, etc. can now be used to generate reliable EAs, wave functions and fine-structure energies. Indeed, for unambiguous and definitive meaning of the EAs of multi-electron atoms and the fullerene molecules our anionic BEs can be used in sophisticated theoretical methods to carry out careful investigations such as has been done in [15] for the At atom.

**Author Contributions:**
Z.F. and A.Z.M. are responsible for the conceptualization, methodology, investigation, formal analysis and writing of the original draft, as well as rewriting and editing. A.Z.M. is also responsible for securing the funding for the research. All authors have read and agreed to the published version of the manuscript.

**Acknowledgments:**
Research was supported by the U.S. DOE, Division of Chemical Sciences, Geosciences and Biosciences, Office of Basic Energy Sciences, Office of Energy Research, Grant: DE-FG02-97ER14743. The computing facilities of National Energy Research Scientific Computing Center, also funded by U.S. DOE are greatly appreciated.

**Conflicts of Interest:** The authors declare no conflict of interest or state.